\documentclass[a4paper,12pt]{article}

\usepackage{amsmath}
\usepackage{amssymb}
\usepackage[dvips]{graphicx}
\usepackage[dvips]{psfrag}
\usepackage{bm}

\newcommand{\be}{\begin{equation}}
\newcommand{\ee}{\end{equation}}
\newcommand{\bea}{\begin{eqnarray}}
\newcommand{\eea}{\end{eqnarray}}
\newcommand{\nn}{\nonumber}
\newcommand{\tr}{{\rm tr}}
\newcommand{\Tr}{\mbox{Tr}}
\newcommand{\Det}{\mbox{Det}}
\newcommand{\cN}{{\cal N}}
\newcommand{\cO}{{\cal O}}
\def\({\left(}
\def\){\right)}

\begin{document}
\thispagestyle{empty} \addtocounter{page}{-1}
\begin{flushright}
SU-HET-03-2014 \\ OIQP-14-07
\end{flushright} 
\vspace*{1cm}

\begin{center}
{\large \bf Tracy-Widom distribution as instanton sum of \\ 2D IIA superstrings }\\
\vspace*{2cm}
Shinsuke M. Nishigaki$^*$ and Fumihiko Sugino$^\dagger$\\
\vskip0.7cm
{}$^*${\it Graduate School of Science and Engineering}\\
\vspace*{1mm}
{\it Shimane University, Matsue 690-8504, Japan}\\
\vspace*{0.2cm}
{\tt mochizuki@riko.shimane-u.ac.jp}\\
\vskip0.4cm
{}$^\dagger${\it Okayama Institute for Quantum Physics} \\
\vspace*{1mm}
{\it Kyoyama 1-9-1, Kita-ku, Okayama 700-0015, Japan}\\
\vspace*{0.2cm}
{\tt fumihiko\_sugino@pref.okayama.lg.jp}\\
\end{center}
\vskip2cm
\centerline{\bf Abstract}
\vspace*{0.3cm}
{\small 
We present
an analytic expression of the nonperturbative free energy of a double-well supersymmetric 
matrix model in its double scaling limit, 
which corresponds to two-dimensional type IIA superstring theory on a nontrivial Ramond-Ramond 
background. 
To this end we draw upon 
the wisdom of random matrix theory
developed by Tracy and Widom,
which expresses the largest eigenvalue distribution of unitary ensembles
in terms of a Painlev\'{e} II transcendent. 
Regularity of the result at any value of the string coupling constant shows that 
the third-order phase transition between a 
supersymmetry-preserving phase and a 
supersymmetry-broken phase, 
previously found at the planar level, 
becomes a smooth crossover in the double scaling limit.
Accordingly, the supersymmetry is always broken spontaneously
as its order parameter stays nonzero 
for the whole region of the coupling constant. 
Coincidence of the result with  
the unitary one-matrix model 
suggests that 
one-dimensional type 0 string theories partially 
correspond to 
the type IIA superstring theory. 
Our formulation naturally 
allows for introduction of 
an instanton chemical potential,
and reveals the presence of a novel phase transition, possibly interpreted as condensation of instantons. 
}

\newpage
\section{Introduction}
Spontaneous supersymmetry (SUSY) breaking in superstring theory is one of crucial
phenomena for superstrings to describe our real world. 
Although various matrix models have been investigated as nonperturbative formulations of 
superstring/M theory~\cite{BFSS,IKKT,MatrixStringMotl,MatrixStringDVV,Maldacena:1997re,Itzhaki:1998dd}, 
it is still difficult to elucidate whether 
these models do 
break SUSY and derive our four-dimensional world. 
In this situation, a simple double-well SUSY matrix model had been recently considered 
in~\cite{Kuroki:2009yg,Kuroki:2010au}, 
and its connection to two-dimensional type IIA superstring theory~\cite{Ita:2005ne} 
on a nontrivial Ramond-Ramond background had been explored from the viewpoint of 
symmetries~\cite{Kuroki:2012nt} and from direct comparison of scattering amplitudes 
at the tree and one-loop orders~\cite{Kuroki:2013qpa}. 
Interestingly, in a double scaling limit that 
realizes the type IIA superstring theory, 
instanton effects of the matrix model survive and break the SUSY spontaneously~\cite{Endres:2013sda}. 
This suggests that the corresponding type IIA superstring theory nonperturbatively breaks its 
target-space SUSY. 
Further investigation along this direction is expected to give insights to nonperturbative SUSY breaking 
in realistic superstring theory.  

In this paper, the nonperturbative computation of the free energy of the SUSY matrix model is completed by
drawing upon the result of Tracy and Widom \cite{Tracy:1994a,Tracy:1994b} on the distribution of
the largest eigenvalue in random matrix theory~\footnote{
Besides 
those quoted in the main text,
the Tracy-Widom distributions at Dyson indices $\beta=2,1,4$ have 
appeared repeatedly 
in the disguise of various
combinatorial and statistical problems
(see \cite{Tracy:2002,Forrester:2012,MS2014} for reviews), 
e.g.~as a distribution of the length of the longest increasing subsequence 
in random permutations \cite{BDJ}, 
as a distribution of 
particles in the  
asymmetric simple exclusion process 
\cite{Johansson,TW2009}, 
and as a one-dimensional
surface growth process in the Karder-Parisi-Zhang universality class \cite{PS2000,Takeuchi,Corwin}. 
}. 
Consequently we shall 
find that the full nonperturbative 
free energy is expressed in terms of a Painlev\'e II transcendent, 
in coincidence with the unitary one-matrix model~\cite{Periwal:1990gf}.
It suggests correspondence between two-dimensional $U(N)$ gauge theory 
and some sector of the two-dimensional IIA superstring theory, 
as well as partial equivalence of the IIA superstrings to one-dimensional type 0 strings. 
The expression is regular for the whole region of the coupling constant, 
and allows expansions in both regions of weak and strong string coupling constants. 
In particular, the third-order phase transition between the SUSY phase and the SUSY-broken 
phase previously 
found in a simple large-$N$ limit (planar limit) disappears in the double scaling limit. 
As a bonus of our method, the free energy or the partition function is naturally generalized by introducing  
instanton fugacity $\xi$. The original free energy or partition function is reproduced as $\xi\to 1$. 

This paper is organized as follows. Our SUSY matrix model is briefly reviewed in the next section, and 
relevant random matrix techniques are summarized in section~\ref{sec:GUE}. 
By combining contents in the above two sections, we present the nonperturbative free energy in 
section~\ref{sec:free_energy}. In section~\ref{sec:condensation}, the generalized free energy is shown to 
exhibit a phase transition due to condensation of instantons 
at an arbitrarily small string coupling constant. 
Section~\ref{sec:discussions} is devoted to summarize the results obtained so far and present 
some of future directions.  
In appendix~\ref{app:TW}, we present some technical steps to the result of Tracy and Widom 
relevant to the text. 

\section{SUSY double-well matrix model}
\label{sec:SUSYMM}
The SUSY double-well matrix model is defined by the 
zero-dimensional reduction of a Wess-Zumino type action 
with superpotential $W(\phi)=\frac13\phi^3-\mu^2\phi$: 
\begin{equation} 
S = N {\rm tr} \left[\frac12 B^2 +iB(\phi^2-\mu^2) +\bar\psi (\phi\psi+\psi\phi)\right],  
\label{S}
\end{equation}
where $B$ and $\phi$ are $N\times N$ hermitian matrices, and $\psi$ and $\bar\psi$ are $N\times N$ 
Grassmann-odd matrices. $S$ is invariant under SUSY transformations generated by $Q$ and $\bar{Q}$: 
\begin{eqnarray}
 & & Q\phi =\psi, \quad Q\psi=0, \quad Q\bar{\psi} =-iB, \quad QB=0, 
\label{QSUSY} 
\\
 & & \bar{Q} \phi = -\bar{\psi}, \quad \bar{Q}\bar{\psi} = 0, \quad 
\bar{Q} \psi = -iB, \quad \bar{Q} B = 0,  
\label{QbarSUSY}
\end{eqnarray}
which leads to the nilpotency: $Q^2=\bar{Q}^2 = \{ Q, \bar{Q}\}=0$. 
In the planar limit (large-$N$ limit with $\mu^2$ fixed), 
the theory has 
two phases -- (I) SUSY phase for $\mu^2>2$ 
and (II) 
SUSY-broken phase for $\mu^2<2$. The phase (I) has infinitely degenerate minima parametrized by 
filling fraction $(\nu_+, \nu_-)$~\footnote{
$\nu_\pm$ are nonnegative fractional numbers such that $\nu_++\nu_-=1$,
corresponding to 
$\nu_+ N$ ($\nu_- N$) eigenvalues of $\phi$
located around the minimum $x=+\mu$ ($x=-\mu$) 
of the double-well potential $V(x)=\frac12 (x^2-\mu^2)^2$.
}, 
and transition between the phases (I) and (II) is of the third order~\cite{Kuroki:2009yg,Kuroki:2010au}.    
As discussed in~\cite{Kuroki:2012nt,Kuroki:2013qpa,Endres:2013sda}, 
various correlation functions of 
the two-dimensional type IIA superstring theory
compactified on $\mathbb{R}\times S^1$ at the selfdual radius,
with the string coupling $g_s$ and the Liouville coupling $\omega$ 
(multiplied by the tachyon operator),
coincide with their counterparts in this matrix model through the identification
$g_s=N^{-1}$ and $4\omega=\mu^2-2$, in 
the double scaling limit 
\be
N\to \infty \quad \mbox{and} \quad \mu^2\to 2+0 \quad \mbox{with}\quad  
s=N^{2/3}(\mu^2-2)
\,=g_s^{-2/3} \cdot 4\omega
\quad  \mbox{fixed}
\label{dsl}
\ee
{}from the phase (I).
Thus, the weakly 
and strongly coupled regions of the IIA superstrings correspond to $s\gg 1$ and $0<s\ll 1$, 
respectively. 
The strength of the Ramond-Ramond background is expressed in terms of $\nu_+-\nu_-$.
After integrating out the auxiliary field $B$ and the fermionic fields $\psi$ and $\bar{\psi}$, 
the partition function of the matrix model can be recast as integrals with respect to 
$N$ eigenvalues
$\{\lambda\}=\{\lambda_1,\ldots,\lambda_N\}$
of $\phi$: 
\begin{equation}
Z(\mu^2) = \tilde{C}_N \int_{-\infty}^\infty \prod_{i=1}^N 
\(d\lambda_i\,e^{-NV(\lambda_i)}\)
\triangle_N(\{\lambda\})^2
\prod_{j,k=1}^N(\lambda_j+\lambda_k), 
\label{Zeigen}
\end{equation}
where $\triangle_N(\{\lambda\}):= \prod_{i>j}^N(\lambda_i-\lambda_j)$, 
$V(\lambda)= \frac12(\lambda^2-\mu^2)^2$,
and $1/\tilde{C}_N =(2\pi)^{\frac{N}{2}}N^{-\frac{N^2}{2}}\prod_{k=0}^Nk!$.
The integration region of each eigenvalue is divided into the positive and negative real axes, and 
the partition function associated with the filling fraction $(\nu_+, \nu_-)$, 
denoted by $Z_{(\nu_+,\nu_-)}(\mu^2)$, 
is defined by integrations over the positive real axis for the first $\nu_+N$ eigenvalues and 
over the negative real axis for the remaining $\nu_-N$. 
Then, it is easy to see the relation 
$Z_{(\nu_+,\nu_-)}(\mu^2) = (-1)^{\nu_-N}\,Z_{(1,0)}(\mu^2)$, where 
\begin{equation}
Z_{(1,0)} (\mu^2) = \tilde{C}_N \int_{0}^\infty \prod_{i=1}^N 
\(d\lambda_i\,e^{-NV(\lambda_i)}\)
\triangle_N(\{\lambda\})^2
\prod_{j,k=1}^N(\lambda_j+\lambda_k). 
\label{Z10}
\end{equation}
The total partition function with a regularization parameter $\alpha$ is defined by 
\be
Z_\alpha(\mu^2) = \sum_{\nu_-N=0}^N\frac{N!}{(\nu_+N)!(\nu_-N)!} \,e^{-i\alpha\nu_-N}
Z_{(\nu_+, \nu_-)}(\mu^2)  = (1-e^{-i\alpha})^N\,Z_{(1,0)}(\mu^2). 
\label{Z_alpha}
\ee
The one-point function $\langle \frac{1}{N}{\rm tr} (iB)\rangle_\alpha$ normalized by 
$Z_\alpha(\mu^2)$ 
coincides with $\langle \frac{1}{N}{\rm tr} (\phi^2-\mu^2)\rangle^{(1,0)}$ normalized by 
$Z_{(1,0)}(\mu^2)$.  
This is well-defined in the limit $\alpha\to 0$ and serves as an order parameter of spontaneous SUSY 
breaking. 

The partition function in the $(1,0)$ sector 
(\ref{Z10})
can be cast in an alternative form by a change of variables $\lambda_i\mapsto x_i=\mu^2-\lambda_i^2$:
\begin{equation}
Z_{(1,0)}(\mu^2)=
\tilde{C}_N \int^{\mu^2}_{-\infty} \prod_{i=1}^N \(dx_i\,e^{-\frac{N}{2} x_i^2}\)
\triangle_N(\{x\})^2
\label{Z10alt}
\end{equation}
Using this expression, one- and two-instanton effects to the one-point function 
and 
$Z_{(1,0)}(\mu^2)$ 
are analytically 
obtained
in \cite{Endres:2013sda}, from which spontaneous breaking of SUSY by instantons is concluded. 
Full nonperturbative contributions are also numerically computed up to 
$N=10^6$, 
and these results are extrapolated to $N=\infty$.
One of the aims of this article is to 
present an {\em analytic} form of full nonperturbative contributions to 
$Z_{(1,0)}(\mu^2)$ 
by recalling {results} in 
random matrix theory.

\section{Gap probability of GUE}
\label{sec:GUE}
Here we collect some basic facts related to
the celebrated result of Tracy and Widom \cite{Tracy:1994a,Tracy:1994b} for completeness.
For an ensemble of sets of $N$ real numbers $\{x\}=\{x_1,\ldots,x_N\}$, 
we consider a joint probability distribution (j.p.d.)\ $P(\{x\})$ 
which is totally 
symmetric under the exchange of any two entries and normalized by 
$\int_{\mathbb{R}}\prod_{i=1}^N dx_i\,P(\{x\})=1$. 
Let us also introduce a function 
associated with an interval $I\subset\mathbb{R}$ by 
\begin{equation}
\tau(\xi;I):=
\int_{\mathbb{R}}\prod_{i=1}^N dx_i(1-\xi \chi_I(x_i)) \,P(\{x\}). 
\label{tau_def}
\end{equation}
Here the characteristic function of $I$ is denoted by $\chi_I(\cdot)$, 
i.e.~$\chi_I(x)=1$ for $x\in I$, and $\chi_I(x)=0$ otherwise. 
In power series expansion of (\ref{tau_def}) with respect to $(-\xi)$, the coefficient of $(-\xi)^k$ 
represents a probability 
in which any $k$ elements of $\{x\}$ are in $I$ and the remaining $(N-k)$ unrestricted 
 (namely, at least $k$ elements are in $I$). 
On the other hand, in expansion with respect to $(1-\xi)$, the coefficient of $(1-\xi)^k$ gives a 
probability 
of exactly $k$ elements belonging in $I$, due to 
$1-\xi\chi_I(x)=\chi_{\mathbb{R}\backslash I}(x)+(1-\xi)\chi_I(x)$. 
These are expressed by the formula: 
\begin{equation}
\tau(\xi;I)
=1+\sum_{k=1}^N \frac{(-\xi)^k}{k!} \int_I dx_1\cdots dx_k R_k(x_1,\ldots,x_k)
=\sum_{k=0}^N (1-\xi)^k E_k(I),
\label{tau}
\end{equation}
where 
\be
R_k(x_1,\ldots,x_k)=\frac{N!}{(N-k)!}\int_{\mathbb{R}}dx_{k+1}\cdots dx_N\,P(\{x\})
\ee 
is the $k$-point correlation function, and 
\be
E_k(I)=\({N\atop k}\)
\int_{I}dx_{1}\cdots dx_k
\int_{\mathbb{R}\backslash I}dx_{k+1}\cdots dx_N\,P(\{x\})
\ee 
is the probability distribution of $k$ elements exclusively in $I$. 
In particular, at $\xi=1$ it is equal to the `gap probability' that the all $x_i$'s 
lie outside the interval $I$,
\begin{equation}
\tau(1;I)=E_0(I)=
\int_{\mathbb{R}\backslash I}\prod_{i=1}^N dx_i \,P(\{x\}).
\label{E0}
\end{equation}

\subsection{Hermitian random matrices}
For an ensemble of $N\times N$ Hermitian random matrices $M$ defined by the partition function of 
the one-matrix model 
\be
Z_{\rm 1MM}
=\int d^{N^2}M\,e^{-\tr \,U(M)} = \tilde{C}_N\int_{\mathbb{R}} \prod_{i=1}^N \(dx_i\,e^{-U(x_i)}\)
\triangle_N(\{x\})^2,
\ee
the corresponding j.p.d. is 
\be
P(\{x\}) = \frac{\tilde{C}_N}{Z_{\rm 1MM}}\,
\prod_{i=1}^N e^{- U(x_i)}
\triangle_N(\{x\})^2. 
\label{jpd_M}
\ee
This j.p.d.~and the $k$-point correlation function $R_k(x_1,\ldots,x_k)$ are 
known to be expressed as a determinant
\begin{equation}
P(\{x\})=
\det [K(x_i, x_j)]_{i,j=1}^N,
\ \ \ 
R_k(x_1,\ldots,x_k)=
\det [K(x_i, x_j)]_{i,j=1}^k
\label{detprocess}
\end{equation}
consisting of a kernel 
\bea
K(x,{y}) & = & e^{-\frac12 (U(x)+U({y}))}\sum_{n=0}^{N-1}\frac{1}{h_n}\,p_n(x)p_n({y}) \nn \\
 & = & e^{-\frac12 (U(x)+U({y}))}\,\frac{1}{h_{N-1}}\,\frac{p_N(x)p_{N-1}({y})-p_{N-1}(x)p_N({y})}{x-{y}}. 
\label{K}
\eea
Here 
$\{p_n(x)\}_{n=0,1,2,\cdots}$ are monic 
polynomials of the degree $n$,
orthogonalized with respect to the measure $e^{-U(x)}dx$: 
\be
\int_{\mathbb{R}} dx\,e^{-U(x)}\,p_n(x)\, p_m(x) = h_n \delta_{nm}. 
\ee 
Furthermore, in terms of the orthonormal functions 
\be
\psi_{n}(x):= \frac{1}{h_n^{1/2}} \,e^{-\frac12 U(x)}\,p_n(x), 
\label{wavefunction}
\ee
the kernel can be cast into a concise form: 
\be
K(x,{y})=\sum_{n=0}^{N-1}\psi_n(x)\,\psi_n({y}).
\label{K_psi}
\ee

Let $\hat{K}|_I$ be an integration operator 
associated with the kernel $K(x,{y})\,\chi_I({y})$ acting on the space of $L^2$ functions on $\mathbb{R}$. 
Although we would like to consider the kernel on the functional space on $I$, it is convenient to treat it 
as an operator on $\mathbb{R}$ by putting the characteristic function~\cite{Tracy:1994a}.  
$\Det$ and $\Tr$ represent the functional determinant and trace over this space, respectively. 
By noting 
\be
\Tr\,
(\hat{K}|_I)^k
=\int_I dx_1\cdots dx_k\,K(x_1, x_2)\,K(x_2, x_3)\cdots K(x_k, x_1),
\ee 
we can see that the Fredholm determinant $\Det(1-\xi \hat{K}|_I)$ has an expansion 
\bea
\Det(1-\xi \hat{K}|_I) & = & \exp\left[\Tr \log (1-\xi \hat{K}|_I)\right] 
= \exp\left[-\sum_{k=1}^\infty\frac{\xi^k}{k}\,\Tr\,(\hat{K}|_I)^k \right] 
\nn \\
& = & 1+\sum_{k=1}^\infty \frac{(-\xi)^k}{k!}\int_I dx_1\cdots dx_k\,\det [K(x_i, x_j)]_{i,j=1}^k. 
\label{Fredholm}
\eea
Here the $k\times k$ matrix $(K(x_i,x_j))_{i,j=1,\cdots, k}$ is a Gram matrix composed by 
the $N$-dimensional real vectors $\vec{\Psi}(x_1), \dots, \vec{\Psi}(x_k)$ with 
$\vec{\Psi}(x)=(\psi_0(x), \cdots, \psi_{N-1}(x))^T\in \mathbb{R}^N$. 
For $k>N$, since the vectors cannot be linearly independent, the Gram determinant vanishes. 
Thus, the infinite series in 
the r.h.s.~of (\ref{Fredholm}) terminates at $k=N$ and coincides with (\ref{tau}). 
This proves the identity 
\be
\tau(\xi;I)={\rm Det}(1-\xi \hat{K}|_I).  
\ee  

\subsection{GUE and soft edge scaling limit}
Now we concentrate on the Gaussian Unitary Ensemble (GUE) defined by the j.p.d. (\ref{jpd_M}) with 
the harmonic oscillator potential $U(x) = \frac{N}2 x^2$, 
for which 
the orthogonal polynomials coincide with 
the Hermite polynomials: 
\be
p_n(x)=\frac{1}{(2N)^{n/2}}H_n\left(\sqrt{\frac{N}{2}}\,x\right) \qquad \mbox{with} \qquad 
H_n(x)=(-1)^n\, e^{x^2}\frac{d^n}{dx^n}e^{-x^2},
\ee 
and the orthonormal functions (\ref{wavefunction}) 
become the wave functions of a particle under a one-dimensional harmonic oscillator potential. 
In a simple large-$N$ limit (planar limit), the eigenvalue density becomes 
\be
\bar{\rho}(x):=
\lim_{N\to \infty}
\frac1N K(x,x)
=\frac{1}{2\pi}\,\sqrt{4-x^2}.
\label{rhobar}
\ee
Let us consider 
another 
large-$N$ limit with $s=N^{2/3}(x-2)$ fixed (the soft-edge scaling limit) which  
unfolds the spectrum near the edge ($x=2$) of the eigenvalue density~(\ref{rhobar}). 
Note that because the edge is nothing but 
one of the classical turning points
of the harmonic oscillator,
the corresponding kernel (the Hermite kernel) $K$ in (\ref{K}) reduces to the Airy kernel:
\be
\lim_{N\to\infty} N^{-2/3}{K}(2+N^{-2/3}s, 2+N^{-2/3}t)
=\frac{{\rm Ai}(s){\rm Ai}'(t)-{\rm Ai}'(s){\rm Ai}(t)}{s-t} 
=:
K_{\rm Ai}(s,t), 
\label{KAi}
\ee
which can be explicitly checked by using the formula~\cite{szego}~\footnote{
For an alternative 
derivation of (\ref{hermite_asymp}), see for example Appendix C in~\cite{Endres:2013sda}.} 
\be
e^{-x^2/2}H_n(x) = \pi^{\frac14} 2^{\frac{n}{2}+\frac14} n^{-\frac{1}{12}}\sqrt{n!}\left[{\rm Ai}(s) 
+\cO(n^{-2/3})\right] 
\label{hermite_asymp}
\ee 
for large $n$ with 
\be
x=\sqrt{2n+1}+\frac{s}{\sqrt{2}\,n^{1/6}}. 
\label{x_s}
\ee
Setting $I=[2+N^{-2/3}s,\infty)$, the scaling limit of $\tau(\xi;I)$ is thus given by
the Fredholm determinant of the Airy kernel,
${\rm Det}(1-\xi \hat{K}_{\rm Ai}|_{[s,\infty)})$.
Tracy and Widom have shown that this quantity 
is expressed as
\cite{Tracy:1994a}:
\begin{equation}
F(\xi,s):=
-\log {\rm Det}(1-\xi \hat{K}_{\rm Ai}|_{[s,\infty)})
=\int_s^\infty (t-s) q(t)^2dt.
\label{TW}
\end{equation}
Here, $q(s)$ 
is a solution to 
a Painlev\'e II differential equation: 
\begin{equation}
q''(s) =s\,q(s)+2q(s)^3, 
\label{PII}
\end{equation}
and is uniquely specified by the boundary condition 
\be
q(s)\sim \sqrt{\xi}{\rm Ai}(s)\ \ (s\to+\infty).
\label{PII_BC}
\ee
In appendix~\ref{app:TW}, we summarize technical points in the derivation of (\ref{TW})-(\ref{PII_BC}). 
{}From the above follows the `specific heat' 
\begin{equation}
\partial_s^2 F(\xi,s)=q(s)^2\sim \xi{\rm Ai}(s)^2.
\label{specificheat}
\end{equation}
Due to (\ref{E0}), the distribution of the (scaled) largest eigenvalue is given by
${\partial_s e^{-F(1,s)}}$. 

It is known that $\tau(\xi;I)=\Det(1-\xi \hat{K}|_I)$ in general is 
a $\tau$ function for the Toda lattice hierarchy associated with
a Painlev\'{e} system. 
In our case, 
$\tau(\xi;I)$ for the Airy kernel
is the one associated with Painlev\'{e} II \cite{Forrester:2001}. 
For a derivation of (\ref{TW})-(\ref{PII_BC}) based on the $\tau$-function theory, see 
the above reference. 

Before closing this section, we comment on a spectrum of the kernel 
{(\ref{K_psi})} or 
its scaling limit~(\ref{KAi}). 
The kernel (\ref{K_psi}) is a projection operator 
acting on $L^2$ functions on $\mathbb{R}$, 
so that 
every 
eigenvalue of $\hat{K}|_{\mathbb{R}}$ is either 0 or 1. 
However, 
considered as an operator acting on $L^2$ functions on an interval $I\subset \mathbb{R}$, 
eigenvalues 
$\{\Lambda_n\}={\rm Spec}(\hat{K}|_I)$ 
are distributed between 0 and 1 in general. 
For the eigenvalue $\Lambda_n$ and the corresponding normalized eigenfunction $f_n(x)$, 
the aforementioned 
upper and lower bounds can be seen from  
\be
\Lambda_n = \int_I dx\,d{y}\,f_n(x)\,K(x,{y})\,f_n({y}) 
=\sum_{m=0}^{N-1}\left(\int_I dx\,f_n(x)\psi_m(x)\right)^2\geq 0 
\label{lowerbound}
\ee
and 
\be
1-\Lambda_n = \int_I dx\, d{y}\,f_n(x)\left(\delta(x-{y})-K(x,{y})\right)f_n({y}) 
=\sum_{m=N}^{\infty}\left(\int_I dx\,f_n(x)\psi_m(x)\right)^2\geq 0.  
\label{upperbound}
\ee
These bounds 
remain valid for the Airy kernel after taking the soft edge scaling limit. 

\section{Free energy and instanton sum}
\label{sec:free_energy}
Our prime 
`observation' is that the partition function of the SUSY double-well matrix model (\ref{Z10alt})
is identical to the gap probability of GUE (\ref{E0}) for $I=[\mu^2,\infty)$, already at finite $N$.
Accordingly, the double-scaling limit in the former (\ref{dsl}) 
is just the soft-edge scaling limit in the latter,
given by (\ref{TW}), (\ref{PII}) and (\ref{PII_BC}) at $\xi=1$:
\be
F(1, s)=-\lim_{N\to \infty}\log Z_{(1,0)}(2+N^{-2/3}s).
\ee 
Notice that the result here is valid for $s<0$ as well as 
for $s\geq 0$. 
Properties of this solution 
to the Painlev\'{e} II equation (\ref{PII}),
called the Hastings-McLeod solution $q_{\rm HM}(s)$ \cite{HM}, are extensively studied in the literature 
(see e.g.~\cite{FIKN}).
Thus we readily have the full nonperturbative free energy of the SUSY double-well matrix model
in the form of (\ref{TW}) with $\xi=1$. 
The free energy is a smooth and positive function of $s$
for the whole range $(-\infty ,\infty)$~\footnote{
Note that $\xi=1$ is the largest value of $\xi$ for these to hold \cite{HM},
as exhibited in the left panel of Fig.~\ref{fig_etacr}.
}.

\subsection{Strong coupling expansion}
With the help of (\ref{PII}), 
$s$-derivatives of $F(1,s)$ at the origin 
are expressed in terms of $q(0)$ and $q'(0)$ of the Hastings-McLeod solution as:
\bea
& & F'(1,0)= -\int^\infty_0 dx \,q(x)^2, \qquad F''(1,0)=q(0)^2, \qquad F'''(1,0)= 2q(0)q'(0), \nn \\
& & F^{(4)}(1,0)= 2q'(0)^2+4q(0)^4, \qquad F^{(5)}(1,0)= 2q(0)^2+24q(0)^3q'(0), \nn \\
& & F^{(6)}(1,0) = 12q(0)q'(0)+72q(0)^2q'(0)^2+48q(0)^6, \nn \\
& & F^{(7)}(1,0)= 20q'(0)^2+64q(0)^4+144q(0)q'(0)^3+576q(0)^5q'(0), \qquad 
\cdots~, 
\eea 
which give a small-$s$ expansion of the free energy. Numerically, we have~\footnote{
This can be obtained either by numerical computation of the Hastings-McLeod solution
or by the Nystr\"om-type method explained in the next section.}
\begin{eqnarray}
F(1,s) & = &
0.0311059853 - 
0.0690913807 s + 
0.0673670913 s^2 - 
0.0361399144 s^3 
\nonumber\\
&&+ 
0.0102959400 s^4 - 
0.000675999388 s^5 - 
0.000468453645 s^6 
\nonumber\\
& & + 
0.0000815342772 s^7-\cdots. 
\label{Fexp}
\end{eqnarray}
Interestingly, the series (\ref{Fexp}) provides 
strong coupling expansion of the IIA superstring theory. 
Smoothness of the free energy shows that
\begin{itemize}
\item
While the third-order phase transition is found in the planar limit for this model~\cite{Kuroki:2010au}, 
it 
turns into a crossover in the double scaling limit and the phases (I) and (II) are smoothly connected without 
any phase transition. 

As its interpretation in the type IIA superstring theory, the planar limit corresponds to extracting 
the string theory at the tree level, where  
the SUSY breaking at the classical level occurring in the phase (II) is distinct from the breaking due to the 
nonperturbative effects in the phase (I). 
However, in the double scaling limit giving a nonperturbative construction of the string theory, 
the difference of the two phases cannot be seen 
in the free energy $F(1,s)$, 
and expressions of the free energy for both regions  
are analytically connected~\footnote{
Since the planar solution for $\mu^2<2$ has a symmetric eigenvalue distribution 
with the support of a single interval~\cite{Kuroki:2010au}, 
physics of the region $s<0$ should connect to that of the region $s>0$ 
with the filling fraction $(1/2, 1/2)$ rather than $(1,0)$. 
However, concerning the free energy the argument in the text is valid, 
because as seen from the relation above eq.~(\ref{Z10}), 
the free energy with the filling fraction $(1/2, 1/2)$ is equal to that with $(1,0)$, i.e. $F(1,s)$, except 
an unimportant additive constant. 
}. 
\item
The above situation is identical with what was 
seen in the unitary one-matrix model 
of two-dimensional $U(N)$ lattice gauge theory \cite{Periwal:1990gf} or 
of one-dimensional type 0 string theories \cite{Klebanov:2003wg}.
The unitary matrix model has two phases in the planar limit, 
which correspond to weakly and strongly coupled regions of the gauge theory, 
respectively. Transition between these phases is
also of the third order~\cite{Gross:1980he,Wadia:1980cp}. 
A double scaling limit of the model (and its generalized versions) was investigated 
by using orthogonal polynomial methods in~\cite{Periwal:1990gf}, where the second derivative of the free energy is 
given in terms 
of the Hastings-McLeod solution.
The functional form of the free energy is essentially the same as our result except the leading planar contribution, which is
smooth across
the two phases, 
i.e. there is no phase transition any longer in the double scaling limit~\footnote{
The same result is obtained in a continuum formulation of the gauge theory
\cite{GrossMatitsyn,FMS2011}.
}.
That issue is discussed in the context of trans-series and resurgence 
in \cite{Marino:2008ya,Schiappa:2013opa}~\footnote{ 
Methods of trans-series and resurgence have been recently investigated in matrix 
models~\cite{Marino:2007te,Pasquetti:2009jg,Aniceto:2011nu} and in quantum field 
theory~\cite{Dunne:2012ae,Basar:2013eka,Cherman:2014ofa}.
}. 
\item
In the double scaling limit of our model, 
aspects of nonperturbative SUSY breaking for the region $s>0$ 
carry over to the region $s<0$ where SUSY is broken at the classical level.   
In fact, the order parameter of spontaneous SUSY breaking
$\left\langle \frac1N \tr (iB)\right\rangle_\alpha$, 
which is proportional to the first $s$-derivative of the free energy, 
also crosses smoothly over from $s>0$ to $s<0$. 
It is worth noting that this realizes analyticity in the spontaneous SUSY breaking 
in spite of the infinite degrees of freedom 
at the large $N$. 
The issue of the analyticity 
is discussed in section~4 of~\cite{Witten:1982df}. 
Although the fact of the SUSY breaking has been observed analytically from the one- or two-instanton 
contributions and numerically as well in~\cite{Endres:2013sda}, 
the region $s>0$ is focused there. 
Thus, our finding of the full nonperturbative 
free energy valid for $s\in (-\infty, \infty)$ 
provides 
a new insight into 
the analytic structure of the 
IIA superstring theory. 
\end{itemize}

\subsection{Weak coupling expansion}
\label{sec:weak}
Asymptotic expansion of the free energy $F(1,s)$ for $s\to \infty$,
which corresponds to 
weak coupling expansion of the IIA 
superstring theory,
can be derived in the following two ways.
The Fredholm expansion in the first line of (\ref{Fredholm}) applied for the Airy kernel (\ref{KAi}) 
decomposes the free energy into a sum of finite-dimensional integrals, 
\bea
F(1,s) & = & \sum_{k\geq 1} F_{k{\rm -inst}}(s),\nn \\
F_{k-{\rm inst}}(s)& := & 
\frac{1}{k}
\int_s^\infty dt_1\cdots dt_k \,K_{\rm Ai}(t_1, t_2)\,K_{\rm Ai}(t_2, t_3)\cdots K_{\rm Ai}(t_k, t_1).
\label{Fkinst}
\eea
In particular, the one-instanton part 
\be
F_{1{\rm -inst}}(s)=\int_s^\infty dt\,K_{\rm Ai}(t,t)=\int_s^\infty dt\({\rm Ai}'(t)^2-t\,{\rm Ai}(t)^2\)
\ee
agrees with eqs.~(5.26), (5.31) of ref.~\cite{Endres:2013sda}
derived directly from the
properties of the Hermite polynomials~\footnote{
The coupling constant $t$ in \cite{Endres:2013sda} corresponds to 
$s/4$ in this paper.
}.
Note that 
asymptotic expansion of the Airy function consists of a single exponential~\cite{Olver}
\begin{equation}
{\rm Ai}(s)\sim \frac{e^{-\frac23 s^{3/2}}}{2\sqrt{\pi} s^{1/4}}\sum_{n=0}^\infty
\frac{(-1)^n (6n)!}{576^n (2n)!(3n)! s^{3n/2}}
\ \ \mbox{for}\ \  s\to \infty 
\label{AiryExp}
\end{equation}
and does not contain subleading exponentials (trans-series). 
$F_{k-{\rm inst}}(s)$ contains 
$2k$-fold products of 
the Airy function, and thus consists of a single exponential 
$e^{-\frac{4k}{3} s^{3/2}}$ times an asymptotic series in $s^{-3/2}$ 
for $s\to \infty$. 
This observation leads to 
identifying $F_{k{\rm -inst}}(s)$ as a $k$-instanton contribution
to the free energy, thereby justifying the notation~\footnote{
It may be possible to obtain the result of $F_{k-{\rm inst}}$ by integrating $k$ eigenvalues in the region of 
$[2, 2+N^{-2/3}s]$ and the remaining $(N-k)$ in the region $[-2,2]$ in (\ref{Z10}) 
as discussed in section 3 of \cite{Endres:2013sda}. 
However, it seems a technically formidable task for general $k$, 
although the 
technique 
of an isomonodromic system~\cite{Witte} 
would manage to 
deal with the cases of small $k$.
}.

Alternatively, one might as well employ 
asymptotic expansion of the Painlev\'{e} II transcendent $q(s)$
for $s\to \infty$ in (\ref{TW}) as presented in \cite{PS,ForBook}.  
Concretely, one substitutes a trans-series
\be
q(s)=
\sqrt{\xi} \sum_{k\geq 0}\xi^k Q_{k}(s) 
\quad (s\to \infty)
\ee
with
\be
Q_{0}
(s)={\rm Ai}(s) \qquad  \mbox{and} \qquad Q_{k}(s)\sim
\frac{e^{-\frac{4k+2}{3} s^{3/2}}}{s^{(6k+1)/4}}\sum_{n=0}^\infty \frac{a^{(k)}_n}{s^{3n/2}}
\ee
into (\ref{PII}) and equates like terms. 
Then, recurrence equations 
determining the coefficients $a^{(k)}_n$ are obtained~\cite{ForBook}. 
Here we list first few terms in each $Q_{k}(s)$:
\begin{eqnarray}
Q_{1}
(s)&\sim&
\frac{e^{-2 s^{3/2}} }{2^5
\pi^{3/2} s^{7/4}}
\left(1-\frac{23}{16 s^{3/2}}+\frac{1493}{512s^3}-\frac{191635}{24576 s^{9/2}}+\cdots\right), \\
Q_{2}
(s)&\sim&
\frac{e^{-\frac{10}{3}s^{3/2}}}{2^9
\pi^{5/2} s^{13/4}}
\left(1-\frac{139}{48 s^{3/2}}+\frac{38005}{4608 s^3}-\frac{17423767}{663552 s^{9/2}}+\cdots\right) , \\
Q_{3}
(s)&\sim&
\frac{e^{-\frac{14}{3} s^{3/2}}}{2^{13}
\pi^{7/2} s^{19/4}} 
\left(1-\frac{209}{48 s^{3/2}}+\frac{72373}{4608 s^3}-\frac{37964645}{663552 s^{9/2}}+\cdots\right), \\
\cdots & &  \nn .  
\end{eqnarray}
After the integration $\int_s^\infty dt\,(t-s)q(t)^2$ of these asymptotics
including (\ref{AiryExp}), 
weak coupling expansion of the free energy is also 
expressed as a trans-series: 
\begin{eqnarray}
F(\xi,s)&=&\sum_{k\geq 1} \xi^k F_{k-{\rm inst}}(s), 
\label{Ftrans}
\\ 
F_{1-{\rm inst}}(s)&\sim&
\frac{e^{-\frac{4}{3} s^{3/2}} }{16 \pi  s^{3/2}}
\left(1-\frac{35}{24 s^{3/2}}+\frac{3745}{1152 s^3}-\frac{805805}{82944 s^{9/2}}+\cdots\right),  \\
F_{2-{\rm inst}}(s)&\sim&
\frac12 \(\frac{e^{-\frac{4}{3} s^{3/2}} }{16 \pi  s^{3/2}}\)^2
\left(1-\frac{35}{12 s^{3/2}}+\frac{619}{72 s^3}-\frac{592117}{20736 s^{9/2}}+\cdots\right), 
\label{F2inst} \\
F_{3-{\rm inst}}(s)&\sim&
\frac13 \(\frac{e^{-\frac{4}{3} s^{3/2}} }{16 \pi  s^{3/2}}\)^3
 \left(1-\frac{35}{8 s^{3/2}}+\frac{2059}{128 s^3}-\frac{184591}{3072 s^{9/2}}+\cdots\right), \\
F_{4-{\rm inst}}(s)&\sim&
\frac14 \(\frac{e^{-\frac{4}{3} s^{3/2}} }{16 \pi  s^{3/2}}\)^4
  \left(1-\frac{35}{6 s^{3/2}}+\frac{3701}{144 s^3}-\frac{1112077}{10368 s^{9/2}}+\cdots\right),
\label{F4inst} \\
\cdots & & \nn .
\end{eqnarray}
Without contribution from perturbative parts, the above asymptotics valid for $s\to\infty$ consists solely of
nonperturbative parts, carrying the instanton action $\frac{4}{3} s^{3/2} \propto N$ 
and expanded in $1/s^{3/2}\propto N^{-1}$.  
It seems plausible 
that the target-space SUSY in the two-dimensional IIA theory is always broken by D-brane like objects.
In view of (\ref{Ftrans})-(\ref{F4inst}), we observe that 
the leading and next-to-leading terms in each $F_{k-{\rm inst}}(s)$ could be resummed in a form 
\be
F(\xi, s)=-\log \left(1-\xi F_{1-{\rm inst}}(s)\right) 
+ \cO\left(\xi^k e^{-\frac{4k}{3}s^{3/2}}s^{-\frac{3k}{2}-3}\right)
\ee
with $k\geq 2$. 
The second term $\cO(\cdots)$ contains contributions from the third or higher terms in $F_{k-{\rm inst}}$ 
for all $k\geq 2$. 
Assuming that this also holds for 
higher-instanton effects, 
the generalized partition function 
(which we call the `grand' partition function)
becomes 
\be
\Xi
(\xi, s):=
e^{-F(\xi,s)}=1-\xi F_{1-{\rm inst}}(s) + \cO\left(\xi^k e^{-\frac{4k}{3}s^{3/2}}s^{-\frac{3k}{2}-3}\right)
\qquad (k\geq 2). 
\label{dilutegas}
\ee
The leading and next-to-leading terms in $F_{k-{\rm inst}}(s)$ represent contributions of $k$ instantons 
and their fluctuations up to the two-loop order. 
Hence, concerning the (grand) partition function (\ref{dilutegas}), 
multi-instanton contributions vanish 
up to this order and start from the three-loop order. 
It is distinct from the dilute gas picture of instantons and suggests significance of interactions 
among instantons. 

\begin{figure}[h]
\centering
\includegraphics[width=78mm]{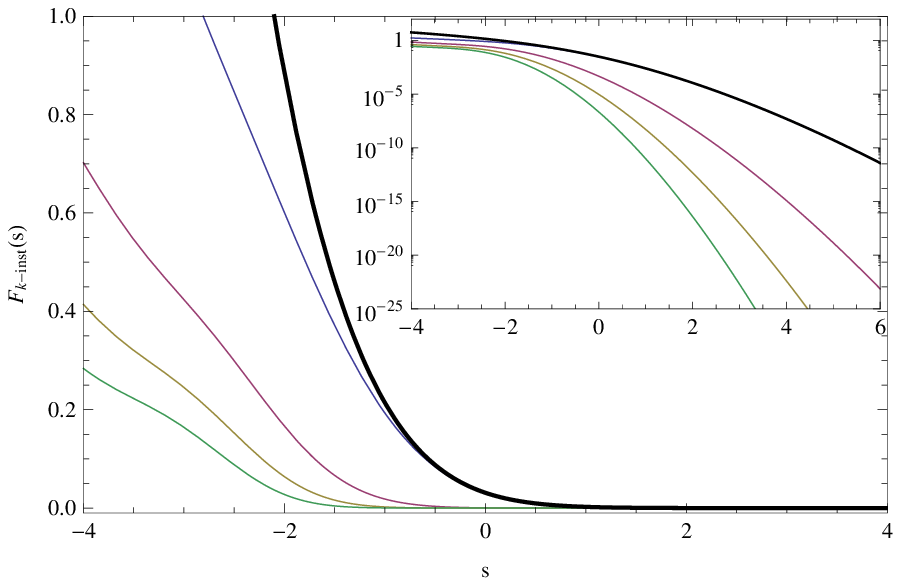}~
\includegraphics[width=78mm]{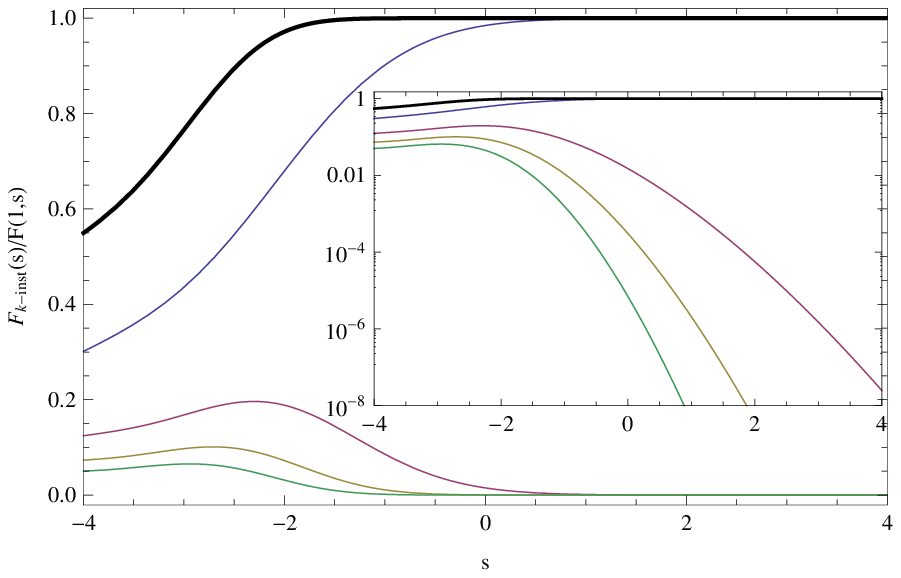}
\caption{Left: 
the free energy $F(1,s)$ (black) and its 1- (blue), 2- (red), 3- (yellow), 4-instanton (green)
parts.\ \ 
Right: relative portions $F_{k-{\rm inst}}(s)/F(1,s)$ 
of $1$- (blue), 2- (red), 3- (yellow), 4-instanton (green)
contributions to the free energy, and the sum of these four (black). 
}
\label{fig_Fk}
\end{figure}

In a technical aspect, 
the above method is considerably easier than asymptotic expansion of 
a closed expression (\ref{Fkinst})
involving $k$-fold integrations. 
The first term of the two-instanton part  
(\ref{F2inst}) 
was previously derived
in \cite{Endres:2013sda}, eq.(6.33).
In Fig.~\ref{fig_Fk} we exhibit  
numerical plots of the free energy and its $k$-instanton parts.
This extends Fig.~4 of the aforementioned reference by including contributions of higher instantons and 
the range of $s<0$.

\subsection{Beyond the strong coupling region}
\label{sec:beyond_weak}
Our identification of the matrix model with 
the two-dimensional IIA superstrings (\ref{dsl}) is limited to the region $s>0$ by construction,
as $s<0$ would formally correspond to a negative Liouville coupling $\omega$ or imaginary string coupling $g_s$.
Nevertheless, the aforementioned smoothness of the free energy as plotted in Fig.~\ref{fig_Fk}
leads us to speculate that the $s<0$ region of the matrix model describes some physical system
whose weak coupling limit is realized as the IIA superstring theory.
As a possible clue in identifying such a system, 
below we exhibit the asymptotic form of the free energy (at $\xi=1$) in the limit $s\equiv -z\to -\infty$.
To this end, we substitute into the Painlev\'{e} II equation a formal trans-series ansatz containing
a single parameter $C$ \cite{Marino:2008ya}:
\begin{equation}
q(z;C)=\sum_{\ell\geq 0} C^\ell q_{\ell}(z),\ \ 
q_{\ell}(z)\sim\frac{e^{-\ell\frac{2\sqrt{2}}{3}z^{3/2}}}{z^{(3\ell-2)/4}}\sum_{n=0}^\infty \frac{b_{n}^{(\ell)}}{z^{3n/2}}
\ \ (z\to\infty) 
\label{formalTS}
\end{equation}
with $q_{0}(z)\sim\sqrt{z/2}+\cdots$, and $b_{0}^{(1)}=2^{-1/4}$ by definition.
By equating like terms as in section~\ref{sec:weak}, 
the `perturbative part' $q_{0}(z)$ is given by \cite{Tracy:1994a}
\begin{equation}
q_{0}(z)\sim
\sqrt{\frac{z}{2}}
\left(
1
-\frac{1}{2^3 z^3}
-\frac{73}{2^7 z^6}
-\frac{10657}{2^{10} z^9}
+\cdots
\right),
\label{ql0}
\end{equation}
and the `nonperturbative parts' $q_{\ell}(z)\ (\ell\geq 1)$ are \cite{Marino:2008ya}
\begin{eqnarray}
q_{1}(z)
&\sim&
\frac{e^{-\frac{2 \sqrt{2}}{3} z^{3/2}}}{2^{1/4} z^{1/4}}
\left(
1
-\frac{17}{2^{9/2} 3 z^{3/2}}
+\frac{1513}{2^{10} 3^2 z^3}
-\frac{850193}{2^{29/2}3^4 z^{9/2}}
+\cdots
\right),\nonumber\\
q_{2}(z)
&\sim&
\frac{e^{-\frac{4 \sqrt{2}}{3} z^{3/2}}}{2z}
\left(1
-\frac{41}{2^{7/2} 3 z^{3/2}}
+\frac{5461}{2^8 3^2 z^3}
-\frac{1734407}{2^{23/2}3^4 z^{9/2}}
+\cdots
\right),\nonumber\\
q_{3}(z)
&\sim&
\frac{ e^{-2 \sqrt{2} z^{3/2}}}{2^{7/4}z^{7/4}}
\left(
1
-\frac{47}{2^{9/2}  z^{3/2}}
+\frac{5285}{2^{10} z^3}
-\frac{1193755}{2^{29/2} 3z^{9/2}}
+\cdots
\right),\nonumber \\
q_{4}(z)
&\sim&
\frac{e^{-\frac{8\sqrt{2} }{3} z^{3/2}}}{2^{5/2}z^{5/2}}
\left(
1
-\frac{25}{ 2^{3/2} 3 z^{3/2}}
+\frac{5011}{2^6 3^2 z^3}
-\frac{1808341}{2^{19/2} 3^4 z^{9/2}}
+\cdots
 \right),
 \nonumber \\
 \cdots& & . \label{ql} 
\end{eqnarray}
Since the positive $z$-axis (i.e.~negative $s$-axis) is 
a Stokes line of the Painlev\'{e} II equation, 
one must perform lateral Borel resummations of the formal series (\ref{formalTS}) by avoiding singularities from above or below, 
which is denoted by $q_{\pm}(z;C)$.
Then the Hastings-McLeod solution is known to be expressed as a median resummation at $C=0$ \cite{Marino:2008ya},
\begin{equation}
q_{\rm HM}(-z)=q_\pm(z;\mp S/2)
=\Re{\rm e} \left[ q_{0,\pm}(z)-\frac{1}{4}S^2 q_{2,\pm}(z)+\frac{5}{16}S^4 q_{4,\pm}(z)+\cdots\right].
\end{equation}
Here $S=-{i}/{\sqrt{2 \pi }}$ is the Stokes constant computed in \cite{FIKN,KKM}, 
and both of the branches give the same result.
Substituting (\ref{ql0}), (\ref{ql}) and integrating $q_{\rm HM}(-z)^2$ twice in $z$, one finally obtains the 
asymptotics of the free energy $F(1,-z)$ for $z\to\infty$:
\begin{eqnarray}
F(1,-z)&\sim&
\frac{z^3}{12}
+\frac{1}{8}\log z
 -\frac{1}{24}\log 2-\zeta'(-1)
 -\frac{3}{2^6 z^3}
 -\frac{63}{2^8 z^6}
+\cdots 
\nonumber\\
&&+
\frac{e^{-\frac{4 \sqrt{2} }{3}z^{3/2}}}{2 \pi  z^{3/2}}
\left(
\frac{1}{2^{11/2}}
-\frac{71}{2^9 3 z^{3/2}}
+\frac{13465}{2^{27/2} 3^2 z^3}
-\frac{5083145}{2^{17}3^4 z^{9/2}}
+\cdots
\right)
\nonumber\\
&&+
\left(\frac{e^{-\frac{4 \sqrt{2} }{3}z^{3/2}}}{2 \pi  z^{3/2}}\right)^2
\left(
\frac{3}{2^{10}}
-\frac{65}{2^{25/2}z^{3/2}}
+\frac{3905}{2^{15} 3 z^3}
-\frac{3132385}{ 2^{39/2} 3^3 z^{9/2}}
+\cdots
\right)
\nonumber\\
&&+\cdots. 
\label{TWasymptotics}
\end{eqnarray}
The integration constant in the above was first conjectured by Tracy and Widom \cite{Tracy:1994a} and later proved true in \cite{DIK08}.
Note that the leading (perturbative) part of the asymptotics is 
an expansion 
in $1/z^{3}$, i.e. each term being proportional to $N^{2-2h}$ with $h\geq 0$,
reminiscent of non-supersymmetric closed strings, 
whereas the nonperturbative parts carrying the instanton action 
$\frac{4\sqrt{2}}{3} z^{3/2} \propto N$ 
are expansions in $1/z^{3/2}\propto N^{-1}$, indicating their open string origin.

We have presented asymptotic behavior of the free energy as $s\to \infty$ in section~\ref{sec:weak} and 
as $s\to -\infty$ here, separately by using trans-series with a single parameter. 
The instanton effects are different for these regions.
For instance, the instanton action in the former (\ref{Ftrans}) is $\frac43 s^{3/2}$, 
while that in the latter (\ref{TWasymptotics}) is 
$\frac{4\sqrt{2}}{3} (-s)^{3/2}$. 
It would be interesting to understand the difference 
from the point of view of resurgence. As discussed in \cite{Schiappa:2013opa}, two-parameter trans-series 
would play a central role in order to perform such a resurgent analysis, which could give an insight into 
the global structure of the free energy for a complex variable $s$.  

\section{Condensation of instantons}
\label{sec:condensation}
We have identified ${\rm Det}(1-\hat{K}_{\rm Ai}|_{[s,\infty)})$ with the double-scaled partition function
of the SUSY double-well matrix model $Z_{(1,0)}(\mu^2)$ at $\mu^2=2+N^{-2/3}s$.
In the 
grand
partition function 
\be
\Xi
(\xi, s)={\rm Det}(1-\xi \hat{K}_{\rm Ai}|_{[s,\infty)}),
\ee
$\xi$ can be regarded as 
fugacity for the matrix model instantons, which should correspond to solitonic objects like D-branes 
in the 
two-dimensional IIA superstring theory. 
Although the instanton fugacity is not explicitly incorporated in the original matrix model 
with the action (\ref{S}), it is 
pleasant surprise that $\xi$ can be naturally introduced into our formulation. 

Now that the parameter space of the model is extended to include an
instanton chemical potential $\eta=\log \xi$ in addition to the original coupling constant $s$,
let us look for a critical line in the $(s, \eta)$-plane.
Note that the partition function is a characteristic `polynomial' of
the Fredholm eigenvalues $\{\Lambda_n(s)\}={\rm Spec}(\hat{K}_{\rm Ai}|_{[s,\infty)})$ as 
\begin{equation}
\Xi
(e^{\eta},s)=
\prod_{n}\(1-e^{\eta} \Lambda_n(s)\), 
\label{Zcp}
\end{equation}
where $1\geq \Lambda_1(s)>\Lambda_2(s)>\cdots\geq 0$ from the argument at the end of section~\ref{sec:GUE}, 
eqs.~(\ref{lowerbound}) and (\ref{upperbound}). 
If one gradually enhances multi-instanton contributions by
turning on a positive  
chemical potential $\eta>0$ at fixed $s$,
the grand partition function vanishes and the corresponding free energy $F(e^\eta, s)$ 
diverges logarithmically whenever
$e^{-\eta}$ approaches 
one of the $\Lambda$'s.
This property could as well be deduced from the expression of the specific heat (\ref{specificheat})
in terms of a Painlev\'{e} II transcendent $q(s)$. Namely, all of its singularities $\{s_n(\eta)\}$
are simple poles that are
movable subject to a change of the boundary condition,
i.e.~the value of 
$\xi=e^{\eta}$ in (\ref{PII_BC}).
This leads to 
$\partial_s^2 F(e^\eta, s)=q^2\propto (s_n(\eta)-s)^{-2}$ and 
$F(e^\eta, s) \propto \log (s_n(\eta)-s)$ 
near any one of the singularities in $s$.
The critical line accessible from the `ordinary' phase $\eta=0$ is dictated by the largest
Fredholm eigenvalue,
\begin{equation}
\eta_{\rm cr}(s)=-\log \Lambda_1(s).
\label{etacr}
\end{equation}
We consider that
this criticality allows an interpretation as a phase transition 
due to condensation of instantons, at least for a sufficiently large positive $s$
where the picture of instantons is valid. 
Subleading eigenvalues $\Lambda_{n\geq 2}(s)$ give
a sequence of singularities, 
but their physical 
or statistical-mechanical significance is unclear as the grand partition function alternates its sign and
becomes negative in the regions 
$-\log \Lambda_{2n-1}(s)< \eta<-\log \Lambda_{2n}(s)$ ($n=1,2,\cdots$). 

\begin{figure}[h]
\centering
\includegraphics[width=80mm]{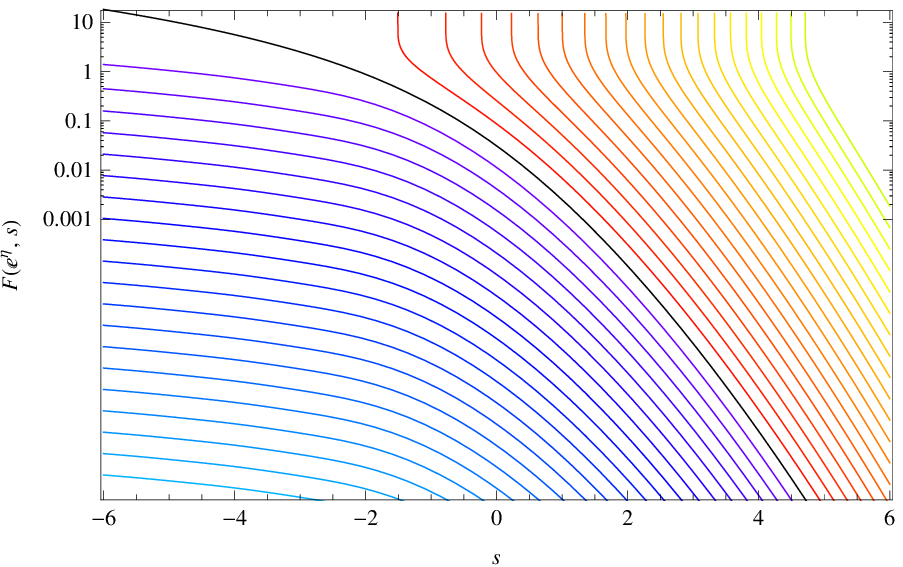}~
\includegraphics[width=78mm]{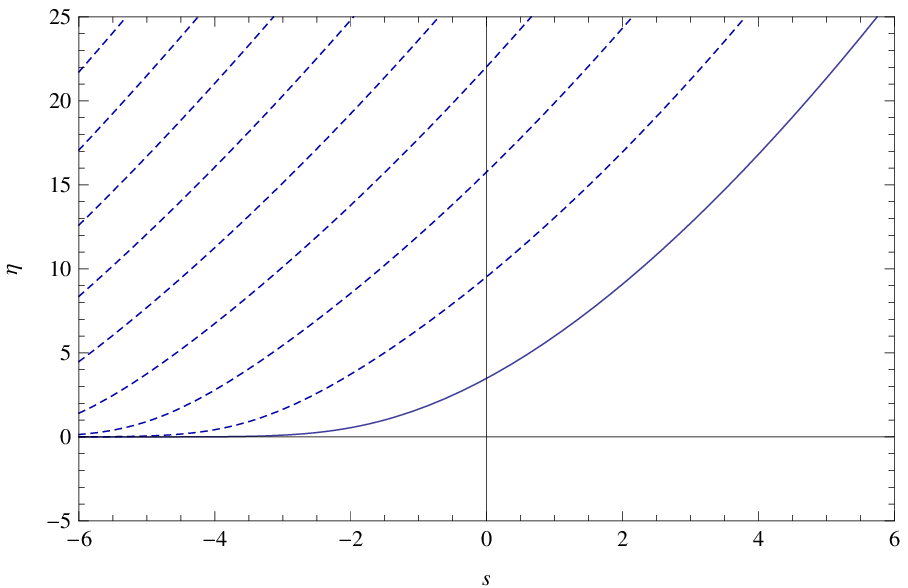}
\caption{Left: the free energy $F(e^{\eta},s)$ for $\eta=-20, -19, \ldots,  -1$ (cyan to purple), 
$\eta=0$ (black), and $\eta=1, 2, \ldots, 20$ (red to green). 
Right: the critical line $\eta=-\log \Lambda_1(s)$ (solid curve) in the ($s,\eta$) plane.
Subleading Fredholm eigenvalues $-\log \Lambda_{2,3,\cdots}(s)$ are also plotted in broken curves.}
\label{fig_etacr}
\end{figure}

As a remark for precise numerical calculation of the spectrum of a trace-class integral operator $\hat{K}|_{I}$,
the so-called Nystr\"{o}m-type method (i.e.\ quadrature approximation) is practically most suited \cite{Bor}.
Namely, after normalizing 
the interval $I$ to $[-1,1]$ by a linear transformation, 
one uses the Gauss quadrature method to discretize 
it into the nodes of the $M$-th order Legendre polynomial $\{x_i\}_{i=1}^M$ such that
$\int_I f(x) dx \simeq \sum_{i=1}^M f(x_i)w_i$.
Here $\{w_i\}_{i=1}^M$ denotes 
appropriate positive weights reflecting the density of the nodes.
Then the integral operator is discretized into an $M\times M$ real symmetric matrix
\be
\hat{K}|_{I}\simeq [K(x_i,x_j)\sqrt{w_i w_j}]_{i,j=1}^M,
\ee
whose eigenvalues can be easily obtained.
When applied to the computation of the Fredholm determinant
${\rm Det}(1-\xi \hat{K}|_{I})\simeq \det[\delta_{ij}-\xi K(x_i,x_j)\sqrt{w_i w_j}]_{i,j=1}^M$,
the discretization error is shown to be 
suppressed as $\cO(e^{-({\rm const.})\,M})$ \cite{Bor}.
For our purpose of computing the Fredholm eigenvalues and determinant for
the Airy kernel (which decreases rapidly for large positive argument(s)) 
in the range $I=[s, \infty)$ with $|s|\lessapprox 1$,
it is sufficient (actually an overkill) 
to truncate the upper range at $s_{\rm max}= 8\sim 10$ and choose $M=100\sim 200$ to
achieve double-precision accuracy.
In Fig.~\ref{fig_etacr} we exhibit plots of 
$F(e^{\eta},s)$ and $\eta_{\rm cr}(s)$ 
computed by this method.
It is evident from the 
plots that for a large positive $s$,
the critical value of $\eta$ approaches 
infinity as
\be
\eta_{\rm cr}(s)\sim \frac43 s^{3/2}+\log(16\pi s^{3/2}),
\ee
in consistency with the first two terms of (\ref{dilutegas}).
This means that even in the weakly coupled region in $s$,
sufficient enhancement of multi-instantons always
drives the system to the phase transition of instanton condensation.

\section{Discussions}
\label{sec:discussions}
We have identified Tracy and Widom's cumulative distribution of the largest eigenvalue of GUE
as the partition function $Z_{(1,0)}(\mu^2)$ 
of the SUSY double-well matrix model describing
two-dimensional IIA superstring theory
on a nontrivial Ramond-Ramond background.
Using this equivalence, strong and weak coupling expansions of the free energy are provided 
in closed forms by a Painlev\'{e} II transcendent. 
Conceptually, the equivalence leads to a novel observation that 
the spontaneous breaking of the target-space SUSY in the IIA superstring theory is realized,
in terms of the quantum mechanics of the eigenvalues of random matrices,
as an exponential tail of the wave function in the classically forbidden domain. 
By interpreting the spectral parameter $\xi$ in the Fredholm determinant of the Airy kernel
as instanton fugacity,
we have identified a phase boundary of a transition due to instanton condensation. 

Some of future subjects worth examining are listed 
below:   
\begin{enumerate}
\item
It is interesting to find an S-dual theory 
which reproduces the strong coupling expansion (\ref{Fexp})
as a perturbation series. Since this 
should be called as a noncritical M-theory, 
it would be helpful to consider a connection with the issue discussed in~\cite{Horava:2005tt}. 
\item 
In this paper, we have focused on the partition function or the free energy of the matrix model. 
In order to make firmer the correspondence between the 
matrix model and the two-dimensional IIA superstrings, 
it is important to proceed computing correlation functions among various 
matrix-model operators 
at higher genera and compare the results with the corresponding IIA string amplitudes. 
For nonperturbative computation beyond the planar level in the matrix model, techniques discussed 
in~\cite{Klebanov:2003wg,Crnkovic:1990mr,DJM,Akemann:1997wi,CK2006} would be useful. 
\item
We have found the equivalence of the free energy of the SUSY matrix model to that of 
the unitary one-matrix model describing the one-dimensional type 0 string theories 
in the double scaling limit. 
It is interesting to investigate whether the equivalence 
persists for quantities other than the free energy. 
To this aim, calculation techniques in 
random matrix theory would be useful to obtain 
correlation functions of various operators in both sides, similarly to the previous subject. 
\item
{}From the viewpoint of 
random matrix theory, 
we list 
three possible extensions of our 
results: 
\begin{itemize} 
\item
We have dealt with the unitary ($\beta=2$) 
ensemble whose matrix variables are complex hermitian. 
For the cases of 
orthogonal and symplectic ($\beta=1, 4$) ensembles 
in which matrix variables are real symmetric 
and quaternion selfdual 
respectively, 
counterparts of the results presented in section~\ref{sec:GUE} have been 
obtained~\cite{Tracy:1995xi,Forrester-Rains,Desrosiers-Forrester}. 
It could be of potential interest to make their interpretations 
in the string theory side, possibly in a relation to non-orientable worldsheets. 
\item
The result in section~\ref{sec:GUE} can be generalized such that the Painlev\'e II 
equation (\ref{PII}) 
contains a parameter $\alpha$~\cite{Claeys}:
\be
q''(s)=s\,q(s)+2q(s)^3+\alpha. 
\label{PII_alpha}
\ee
It reduces to our case in the limit $\alpha\to 0$. 
According to~\cite{Akemann:1997wi,Minahan:1991pv}, 
turning on the parameter $\alpha$ corresponds to introducing 
`quarks' in the matrix models. 
While such quarks generate boundaries in a random surface, 
our interpretation of the matrix model as the IIA superstring theory is 
not based on the random surface picture. 
It is intriguing to pursue what kind of deformations of our matrix model amounts to 
giving (\ref{PII_alpha}) and to find its meaning in the string theory side. 
\item
Multi-critical analogues of the Tracy-Widom distribution 
for $\beta=2$ was studied in \cite{CIK2009,CO2010,Akemann:2012bw} and 
its interpretation as instanton effects in minimal string theory
was presented in~\cite{Atkin:2013mta}. 
It would be interesting to introduce instanton fugacity for such cases and 
discuss instanton condensation as in section~\ref{sec:condensation}. 
\end{itemize}
\item
Alday, Gaiotto and Tachikawa (AGT)~\cite{Alday:2009aq} found correspondence between 
instanton sums of four-dimensional $\cN=2$ SUSY gauge theories 
(the so-called Nekrasov partition functions)~\cite{Nekrasov:2002qd,Nekrasov:2003rj} 
and conformal blocks in two-dimensional Liouville field theory. 
Furthermore, ref.~\cite{Gamayun:2013auu} points out that 
the $\tau$-functions for 
Painlev\'e III, V and VI 
(corresponding to the Fredholm determinants of
e.g.~Bessel, sine and Hermite kernels, respectively) are all 
related to $c=1$ conformal blocks, 
and thus further correspondence is made 
between the instanton sums of $\cN=2$ SUSY gauge theories with $N_f=0,\cdots,4$ 
and the $\tau$-functions 
of the Painlev\'e systems. 
Since we have found the correspondence of the instanton sum of the two-dimensional IIA superstring theory 
to the $\tau$-function for
Painlev\'e II
(corresponding to the Fredholm determinant of Airy kernel), 
they are expected to have an analogous relation to some conformal blocks. 
In addition, existence of the six-dimensional (2,0) theory has been 
argued to lie behind the AGT correspondence. 
Likewise, the similarity in our case will lead to existence of the 
three-dimensional noncritical M-theory mentioned in the first subject.  
\end{enumerate}

\section*{Acknowledgements}
We thank Shinobu Hikami for invitation to the OIST Workshop {\em RMT2013}
that made this collaboration possible. 
We also thank Michael Endres, Peter Forrester and Yuki Sato for helpful communication. 
This work is supported in part
by JSPS Grants-in-Aids for Scientific Research (C) Nos. 25400259 (SMN) and 
25400289 (FS).

\appendix
\section{Derivation of (\ref{TW})-(\ref{PII_BC})}
\label{app:TW}
In this appendix, we present some technical steps relevant to the derivation of (\ref{TW})-(\ref{PII_BC}). 
First, the action of a generic integration operator $\hat{O}$ to a function $f(x)$ on $\mathbb{R}$ 
is expressed by its kernel $O(x,{y})$ as 
\be
\left(\hat{O} f\right)(x) = \int_\mathbb{R} d{y}\,O(x,{y}) f({y})
\qquad \mbox{and}\qquad 
\left(f \hat{O}\right)({y}) = \int_\mathbb{R} dx \,f(x)\, O(x,{y}). 
\ee
Note that for the kernel $K(x,{y})\chi_I({y})$ of the operator $\hat{K}|_I$, 
\bea
\left(\hat{K}|_I f\right)(x) & = & \int_\mathbb{R} d{y}\,K(x,{y})\chi_I({y}) f({y})= \int_I d{y}\,K(x,{y}) f({y}), 
\nn \\ 
\left(f \hat{K}|_I\right)({y}) & = & \int_\mathbb{R} dx \,f(x)\, K(x,{y}) \chi_I({y}). 
\eea

Suppose $K(x,{y})$ takes the form: 
\be
K(x,{y})=\frac{A(x)B({y})-B(x)A({y})}{x-{y}}.  
\ee
For the position operator $\hat{x}$ specified by its kernel $x\,\delta(x-{y})$, 
the kernel of $[\hat{x}, (1- \hat{K}|_I)^{-1}]$ is given by~\footnote{For notational simplicity, 
we absorb $\xi$ into $\hat{K}|_I$ or $K(x,y)$ in this appendix (up to eq.~(\ref{app:R_diffeq3})).
} 
\be
(x-{y})\rho(x,{y})=  
 (Q(x)\,P({y})-P(x)\,Q({y}))\,\chi_I({y}) 
\label{app:x_rho}
\ee
with $\rho(x,{y})$ being the kernel of $(1- \hat{K}|_I)^{-1}$ 
and 
\bea
Q(x) & := & \left((1- \hat{K}|_I)^{-1} A\right)(x) = \int_{\mathbb{R}}d{y}\,\rho(x,{y})\,A({y}), 
\nn
\\
P(x) & := & \left((1- \hat{K}|_I)^{-1} B\right)(x) = \int_{\mathbb{R}}d{y}\,\rho(x,{y})\,B({y}). 
\label{app:def_QP}
\eea
{}From the expansion $(1- \hat{K}|_I)^{-1}= 1+\sum_{n=1}^\infty (\hat{K}|_I)^n$, 
we can see that $\rho(x,y)=0$ for $x\in I$ and $y\notin I$. 
For the resolvent operator 
\be
\hat{R}:=(1- \hat{K}|_I)^{-1} \hat{K}|_I = (1- \hat{K}|_I)^{-1} -1, 
\label{app:R_def}
\ee
its kernel $K(x,{y})$ takes the form 
\be
R(x,y)= \,
\frac{Q(x)\,P(y)-P(x)\,Q({y})}{x-y} \,\chi_I(y)
\label{app:R}
\ee
since the kernel of $[\hat{x}, \hat{R}]$ is nothing but (\ref{app:x_rho}). 
The definition indicates that the diagonal part of the kernel is given by 
the logarithmic derivative of the Fredholm determinant: 
\be
R(a, a)=\frac{d}{da}\,\log\Det(1- \hat{K}|_I). 
\label{app:R2}
\ee
Hereafter we consider the interval $I=[a, \infty)$ 
($a$ will be eventually set to $2+N^{-2/3}s$). 

\subsection{JMMS equations}
In the case of the Hermite kernel of GUE, $A(x)$ and $B(x)$ can be identified with the wave functions: 
$A(x)=\sqrt{\xi} \,\psi_N(x)$ and $B(x)=\sqrt{\xi} \,\psi_{N-1}(x)$. 
Hence they satisfy 
\be
\frac{d}{dx}\begin{pmatrix} A(x) \\ B(x) \end{pmatrix} = 
N\begin{pmatrix} -\frac12 x & 1 \\ -1 & \frac12 x \end{pmatrix} \begin{pmatrix} A(x) \\ B(x) \end{pmatrix}. 
\label{app:d_AB}
\ee
For the derivative operator $\hat{d}$ associated with the kernel $\delta'(x-y)$, the kernel of 
$[\hat{d}, (1- \hat{K}|_I)^{-1}]$ is obtained as 
\be
-\frac{N}{2} \,
(Q(x)\,P(y)+P(x)\,Q(y))\,\chi_I(y) +R(x,a)\,\rho(a, y). 
\label{app:d_rho}
\ee
Use of (\ref{app:def_QP}), (\ref{app:d_AB}) and (\ref{app:d_rho}) leads to 
\bea
\frac{d}{dx}Q(x) & = & -\frac{N}{2}x\,Q(x)+N(1- u)\,P(x)+R(x,a)\,q, \nn \\
\frac{d}{dx}P(x) & = & \frac{N}{2}x\,P(x)-N(1+ w)\,Q(x)+R(x,a)\,p, 
\label{app:d_QP}
\eea
where 
$u = \int_I dx\,Q(x)\,A(x)$, $w = \int_I dx\,P(x)\,B(x)$, $q = Q(a)$ 
and $p = P(a)$. Here and in what follows, quantities at the boundary $x=a$ are 
defined by taking the limit $x\to a +0$, i.e. the limit from the inside of $I$.    
Also, 
\be
\frac{\partial}{\partial a}Q(x)=-R(x,a)\,q, \qquad 
\frac{\partial}{\partial a}P(x)=-R(x,a)\,p 
\label{app:dsig_QP}
\ee
are derived from the fact that the kernel of $
\frac{\partial}{\partial a}(1- \hat{K}|_I)^{-1}$ is 
$-R(x,a)\,\rho(a, y)$.  
Together with this, (\ref{app:d_QP}) gives 
\bea
\frac{dq}{da} & =& \left.\left(\frac{d}{dx}+\frac{\partial}{\partial a}\right)Q(x)\right|_{x=a}
=-\frac{N}{2}a\,q +N(1- u)\,p, \nn \\
\frac{dp}{da} & =& \left.\left(\frac{d}{dx}+\frac{\partial}{\partial a}\right)P(x)\right|_{x=a}
=\frac{N}{2}a\,p -N(1+ w)\,q 
\label{app:d_qp} 
\eea
and
\be
\frac{\partial u}{\partial a} = -q^2, \qquad \frac{\partial w}{\partial a} = -p^2. 
\label{app:d_uw}
\ee
Finally, (\ref{app:R}) at $x=y=a$ is expressed as 
\bea
R(a, a) & = &  
\left.\left[
\Bigl(\frac{d}{dx}Q(x)\Bigr) \,P(x) - 
\Bigl(\frac{d}{dx}P(x)\Bigr) \,Q(x)
\right]\right|_{x=a} \nn \\
& = & N 
\left[
-a \,p\,q +(1- u)\,p^2 +(1+ w)\,q^2
\right]. 
\label{app:R_diag}
\eea
(\ref{app:d_qp}), (\ref{app:d_uw}) and (\ref{app:R_diag}) are the Jimbo-Miwa-M\^{o}ri-Sato (JMMS) 
equations~\cite{Jimbo:1979rt} 
for the half-infinite interval $I$, from which we shall obtain a closed differential equation 
for the diagonal resolvent. 

\subsection{Painlev\'e VI equation for the diagonal resolvent}
{}From the JMMS equations, we find 
\bea
\frac{d}{da}R(a,a) & = & -N\,p\,q, 
\label{app:d_R_diag}
\\
\frac{d}{da}(p\,q) & = & N(1- u)\,p^2-N(1+ w)\,q^2, 
\label{app:d_pq2}\\
\frac{d}{da}(u-w+ uw) & = & (1- u)\,p^2-(1+ w)\,q^2. 
\label{app:d_uw2}
\eea
The last two equations mean that $p\,q$ is equal to $N(u-w+ u w)$ up to an additive $a$-independent 
constant. 
However, the fact that all of $p$, $q$, $u$ and $w$ vanish as $a\to \infty$ determines the constant 
to be nil. 
Namely, 
\be
p\,q=N(u-w+ u w). 
\label{app:pq_uw}
\ee

With the help of (\ref{app:d_R_diag}), (\ref{app:d_pq2}) and the JMMS equations, we have 
\be
\frac{d^3}{da^3}R(a, a) = -N \,\frac{d}{da}\left[
2Np^2q^2+N\,a\{ R(a, a)+N a \,p\,q\} 
-4N^2(1- u)(1+ w)\,p\,q
\right]. 
\ee
After the use of (\ref{app:d_R_diag}) following (\ref{app:pq_uw}), we 
finally obtain a differential equation for $R(a,a)$:
\be
\frac{d^3}{da^3}R(a, a) =-N^2 a\, R(a,a) +N^2(a^2-4)\,\frac{d}{da}R(a, a)
-6\Bigl(\frac{d}{da}R(a, a)\Bigr)^2,
\label{app:R_diffeq}
\ee
which can be transformed to Okamoto's $\sigma$-form 
\cite{Okamoto} of a Painlev\'{e} VI equation. 

\subsection{Soft edge scaling limit and Painlev\'e II equation}
In the soft edge scaling limit $a=2+N^{-2/3}s$, (\ref{KAi}) indicates that 
the diagonal resolvent scales as $R(a, a) = N^{2/3}R(s)$. 
Then, (\ref{app:R_diffeq}) becomes 
\be
R'''(s)=-2R(s)+4s\,R'(s)-6R'(s)^2.
\label{app:R_diffeq2}
\ee 
Integration of (\ref{app:R_diffeq2}) after multiplied by $R''(s)$ leads to 
\be
\frac14\,R''(s)^2=-R(s)\,R'(s) +s\,R'(s)^2-R'(s)^3 . 
\label{app:R_diffeq3}
\ee
Exponential decay of $R(s)$ as $s\to \infty$ is clear from the behavior of the Airy kernel (\ref{KAi}). 
We used it as an initial condition of the integration. 
Setting $R(s)=\int_s^\infty dt\,q(t)^2$, we see that 
the Painlev\'e II equation (\ref{PII}) is obtained from (\ref{app:R_diffeq3}).   
Accordingly, the scaling limit of (\ref{app:R2})
with $\xi$ restored takes the form 
\be
\frac{d}{ds}\log\Det(1-\xi \hat{K}_{\rm Ai}|_{[s,\infty)})= R(s)=
\int_s^\infty dt\,q(t)^2  .
\label{app:Fredholm_diff}
\ee
For $\xi$ small, differentiating (\ref{app:Fredholm_diff}) with respect to $s$ gives  
\be
q(s)^2=\xi\frac{d}{ds}K_{\rm Ai}(s,s) +\cO(\xi^2) = \xi {\rm Ai}(s)^2+\cO(\xi^2)  
\ee 
with use of (\ref{KAi}). 
This yields the boundary condition (\ref{PII_BC}), 
since the $\cO(\xi^2)$ terms consist of higher powers of the Airy function  
and become negligible as $s\to \infty$. 
Finally, (\ref{TW}) follows from (\ref{app:Fredholm_diff}), 
where the integration constant is fixed by 
the small-$\xi$ behavior. 



\begin{thebibliography}{999}
{\small 
\bibitem{BFSS}
T. Banks, W. Fischler, S. H. Shenker and L. Susskind,
Phys.\ Rev.\ D {\bf 55} (1997) 5112 
[{\tt arXiv:hep-th/9610043}].

\bibitem{IKKT}
N. Ishibashi, H. Kawai, Y. Kitazawa and A. Tsuchiya,
Nucl.\ Phys.\ B {\bf 498} (1997)  467
[{\tt arXiv:hep-th/9612115}].

\bibitem{MatrixStringMotl}
L. Motl,
  {\tt arXiv:hep-th/9701025}.
\bibitem{MatrixStringDVV}
 R. Dijkgraaf, E. P. Verlinde and H. L. Verlinde,
  Nucl.\ Phys.\  B {\bf 500} (1997) 43  [{\tt arXiv:hep-th/9703030}].

\bibitem{Maldacena:1997re}
  J. M. Maldacena, 
Adv.\ Theor.\ Math.\ Phys.\  {\bf 2} (1998) 231; 
 [Int.\ J.\ Theor.\ Phys.\  {\bf 38} (1999) 1113] 
[{\tt arXiv:hep-th/9711200}]. 

\bibitem{Itzhaki:1998dd}
  N. Itzhaki, J. M. Maldacena, J. Sonnenschein and S. Yankielowicz,
Phys.\ Rev.\ D {\bf 58} (1998) 046004 
[{\tt arXiv:hep-th/9802042}].

\bibitem{Kuroki:2009yg}
  T. Kuroki and F. Sugino,
Nucl.\ Phys.\ B {\bf 830} (2010) 434  
[{\tt arXiv:0909.3952 [hep-th]}].

\bibitem{Kuroki:2010au} 
  T. Kuroki and F. Sugino,
Nucl.\ Phys.\ B {\bf 844} (2011) 409 
 [{\tt arXiv:1009.6097 [hep-th]}].  

\bibitem{Ita:2005ne}
  H. Ita, H. Nieder and Y. Oz,
JHEP {\bf 0506} (2005) 055 
 [{\tt hep-th/0502187}].  

\bibitem{Kuroki:2012nt}
  T. Kuroki and F. Sugino,
Nucl.\ Phys.\ B {\bf 867} (2013) 448 
[{\tt arXiv:1208.3263 [hep-th]}].  

\bibitem{Kuroki:2013qpa}
  T. Kuroki and F. Sugino,
JHEP {\bf 1403} (2014) 006 
[{\tt arXiv:1306.3561 [hep-th]}].

\bibitem{Endres:2013sda} 
  M. G. Endres, T. Kuroki, F. Sugino and H. Suzuki,
Nucl.\ Phys.\ B {\bf 876} (2013) 758
[{\tt arXiv:1308.3306 [hep-th]}].  

\bibitem{Tracy:1994a}
  C. A. Tracy and H. Widom,
Commun.\ Math.\ Phys.\  {\bf 159} (1994) 151 
 [{\tt hep-th/9211141}].  

\bibitem{Tracy:1994b}
  C. A. Tracy and H. Widom,
Commun.\ Math.\ Phys.\  {\bf 163} (1994) 33 
[{\tt hep-th/9306042}].  

\bibitem{Tracy:2002}
  C. A. Tracy and H. Widom,
in: Proceedings of the ICM. Vol.~1, pp.~587--596 (Beijing, 2002) [{\tt math-ph/0210034}].

\bibitem{Forrester:2012}
P. J. Forrester and N. S. Witte, 
{\tt arXiv:1210.3381 [math-ph]}.
 
\bibitem{MS2014}
S.~N.~Majumdar and G.~Schehr, 
J. Stat. Mech. (2014) P01012
[{\tt arXiv:1311.0580 [cond-mat]}].

\bibitem{BDJ}
J.~Baik, P.~Deift and K.~Johansson, 
J. Amer. Math. Soc. {\bf 12} (1999) 1119
[{\tt math.CO/9810105}].

\bibitem{Johansson}
K.~Johansson,
Commun.~Math.~Phys. {\bf 209} (2000) 437
[{\tt math/9903134 [math.CO]}].

\bibitem{TW2009}
C.~A.~Tracy, and H.~Widom,
Commun.~Math.~Phys. {\bf 290} (2009) 129
[{\tt arXiv:0807.1713 [math.PR]}].
		
\bibitem{PS2000}
M. Pr\"{a}hofer and H. Spohn,
Phys. Rev. Lett. {\bf 84} (2000) 4882 [{\tt cond-mat/9912264}].

\bibitem{Takeuchi}
K. A. Takeuchi, M. Sano, T. Sasamoto and H. Spohn,
Sci. Rep. {\bf 1} (2011) 34
[{\tt arXiv:1108.2118 [cond-mat]}].

\bibitem{Corwin}
I.~Corwin,
Random Matrices: Theory Appl. {\bf 1} (2012) 1130001
[{\tt arXiv:1106.1596 [math.PR]}].

\bibitem{Periwal:1990gf}
  V.~Periwal and D.~Shevitz,
Phys.\ Rev.\ Lett.\  {\bf 64} (1990) 1326;  
%
Nucl.\ Phys.\ B {\bf 344} (1990) 731.  

\bibitem{szego}
G.~Szeg\H{o}, ``Orthogonal Polynomials,'' 4th ed., 
American Mathematical Society 
(Providence, 1975). 

\bibitem{Forrester:2001}
P. J. Forrester and N. S. Witte, 
Commun.\ Math.\ Phys.\  {\bf 219} (2001) 357
[{\tt math-ph/0103025}].  

\bibitem{HM}
S. P. Hastings and J. B. McLeod, 
Arch.\ Rat.\ Mech.\ Anal.\ {\bf 73} (1980) 31.

\bibitem{FIKN}
A.~S.~Fokas, A.~R.~Its, A.~A.~Kapaev and V.~Y.~Novokshenov,
``Painlev\'e Transcendents: The Riemann-Hilbert Approach",
American Mathematical Society (Providence, 2006). 

\bibitem{Klebanov:2003wg}
  I.~R.~Klebanov, J.~M.~Maldacena and N.~Seiberg,
Commun.\ Math.\ Phys.\  {\bf 252} (2004) 275  [{\tt hep-th/0309168}].  

\bibitem{Gross:1980he}
  D.~J.~Gross and E.~Witten,
Phys.\ Rev.\ D {\bf 21} (1980) 446.  

\bibitem{Wadia:1980cp}
  S.~R.~Wadia,
Phys.\ Lett.\ B {\bf 93} (1980) 403.  

\bibitem{GrossMatitsyn}
D.~J.~Gross and A.~Matytsin,
Nucl. Phys. B {\bf 429} (1994) 50
[{\tt hep-th/9404004}].

\bibitem{FMS2011}
P.~J.~Forrester, S.~N.~Majumdar, G.~Schehr, 
Nucl. Phys. B {\bf 844} (2011) 500 
[Erratum ibid. B{\bf 857} (2012) 424] 
[{\tt arXiv:1009.2362 [math-ph]}].

\bibitem{Marino:2008ya}
  M.~Mari\~{n}o,
JHEP {\bf 0812} (2008) 114  [{\tt arXiv:0805.3033 [hep-th]}].  

\bibitem{Schiappa:2013opa}
  R.~Schiappa and R.~Vaz,
Commun.\ Math.\ Phys. {\bf 330} (2014) 655 
[{\tt arXiv:1302.5138 [hep-th]}].

\bibitem{Marino:2007te}
  M.~Mari\~{n}o, R.~Schiappa and M.~Weiss,
  Commun.\ Num.\ Theor.\ Phys.\  {\bf 2} (2008) 349
  [{\tt arXiv:0711.1954 [hep-th]}]; 
%
  J.\ Math.\ Phys.\  {\bf 50} (2009) 052301
  [{\tt arXiv:0809.2619 [hep-th]}].

\bibitem{Pasquetti:2009jg}
  S.~Pasquetti and R.~Schiappa,
  Annales Henri Poincar\'e {\bf 11} (2010) 351
  [{\tt arXiv:0907.4082 [hep-th]}].

\bibitem{Aniceto:2011nu}
  I.~Aniceto, R.~Schiappa and M.~Vonk,
  Commun.\ Num.\ Theor.\ Phys.\  {\bf 6} (2012) 339
  [{\tt arXiv:1106.5922 [hep-th]}].

\bibitem{Dunne:2012ae}
  G.~V.~Dunne and M.~\"{U}nsal,
  JHEP {\bf 1211} (2012) 170
  [{\tt arXiv:1210.2423 [hep-th]}]; 
%
Phys.\ Rev.\ D {\bf 89} (2014) 041701  [{\tt arXiv:1306.4405 [hep-th]}].

\bibitem{Basar:2013eka}
  G.~Basar, G.~V.~Dunne and M.~\"{U}nsal,
JHEP {\bf 10} (2013) 041
 [{\tt arXiv:1308.1108 [hep-th]}].

\bibitem{Cherman:2014ofa}
  A.~Cherman, D.~Dorigoni and M.~\"{U}nsal,
{\tt arXiv:1403.1277 [hep-th]}.  

\bibitem{Witten:1982df}
  E.~Witten,
Nucl.\ Phys.\ B {\bf 202} (1982) 253.  

\bibitem{Olver}
F.~Olver,
``Asymptotics and Special Functions",
Academic Press (London, 1974).

\bibitem{Witte}
N.~S.~Witte, F.~Bornemann and P.~J.~Forrester, 
Nonlinearity {\bf 26} (2013) 1799 [{\tt arXiv:1209.2190 [math.CA]}]. 

\bibitem{PS}
M. Pr\"{a}hofer and H. Spohn,
J. Stat. Phys. {\bf 115} (2004) 255 [{\tt cond-mat/0212519}].

\bibitem{ForBook}
P. J. Forrester, ``Log-Gases and Random Matrices", Princeton Univ. Press (Princeton, 2010).

\bibitem{KKM}
H. Kawai, T. Kuroki and Y. Matsuo,
Nucl.\ Phys.\ B {\bf 711} (2005) 253  [{\tt hep-th/0412004}].

\bibitem{DIK08}
P. Deift, A. Its and I. Krasovsky, 
Commun.\ Math.\ Phys.\ {\bf 278} (2008) 643  [{\tt math/0609451}].

\bibitem{Bor}
F. Bornemann, Math. Comp. {\bf 79} (2010) 871
[{\tt arXiv:0804.2543 [math.NA]}];
Markov Processes Relat. Fields {\bf 16} (2010) 803
[{\tt arXiv:0904.1581 [math.PR]}].

\bibitem{Horava:2005tt}
  P.~Ho\v{r}ava and C.~A.~Keeler,
JHEP {\bf 0707} (2007) 059  [{\tt hep-th/0508024}].  

\bibitem{Crnkovic:1990mr}
  \v{C}.~Crnkovi\'{c} and G.~W.~Moore,
Phys.\ Lett.\ B {\bf 257} (1991) 322.  

\bibitem{DJM}
S.~Dalley, C.~V.~Johnson and T.~Morris,
Nucl.\ Phys.\ B {\bf 368} (1992) 625.

\bibitem{Akemann:1997wi}
  G.~Akemann, P.~H.~Damgaard, U.~Magnea and S.~M.~Nishigaki,
Nucl.\ Phys.\ B {\bf 519} (1998) 682  [{\tt hep-th/9712006}].  

\bibitem{CK2006}
T. Claeys and A.~B.~J.~Kuijlaars,
Comm. Pure Appl. Math. 59 (2006) 1573
[{\tt math-ph/0501074}].

\bibitem{Tracy:1995xi}
  C.~A.~Tracy and H.~Widom,
Commun.\ Math.\ Phys.\  {\bf 177} (1996) 727  [{\tt solv-int/9509007}].  

\bibitem{Forrester-Rains}
P.~J.~Forrester and E.~M.~Rains, 
in: 
(Ed.) P.~M.~Bleher and A.~R.~Its, ``Random matrix models and their applications,'' vol.~40 of 
Mathematical Sciences Research Institute Publications, pp.~171--208, Cambridge Univ. Press
(Cambridge, 2001)
[{\tt solv-int/9907008}].

\bibitem{Desrosiers-Forrester}
P.~Desrosiers and P.~J.~Forrester,
Nonlinearity {\bf 19} (2006) 1643
[{\tt math-ph/0604027}].

\bibitem{Claeys}
T.~Claeys, A.~B.~J.~Kuijlaars and M.~Vanlessen, 
Ann.\ Math.\ {\bf 168} (2008) 601
[{\tt math-ph/0508062}].
 
\bibitem{Minahan:1991pv}
  J.~A.~Minahan,
Phys.\ Lett.\ B {\bf 268} (1991) 29.  

\bibitem{CIK2009}
T.~Claeys, A.~Its and I.~Krasovsky, 
Comm.\ Pure\ Appl.\ Math.\ {\bf 63} (2010) 362 
[{\tt arXiv:0901.2473 [math-ph]}]. 

\bibitem{CO2010}
T.~Claeys and S. Olver,
in: (Ed.) J. Arves\'{u} and G. L\'{o}pez Lagomasino,
``Contemporary Mathematics: Recent Advances in Orthogonal Polynomials, Special Functions, 
and Their Applications'', pp.~83--98, American Mathematical Society
(Providence, 2012) 
[{\tt arXiv:1111.3527 [math-ph]}].

\bibitem{Akemann:2012bw}
  G.~Akemann and M.~R.~Atkin,
J.\ Phys.\ A {\bf 46} (2013) 015202  [{\tt arXiv:1208.3645 [math-ph]}].  

\bibitem{Atkin:2013mta}
  M.~R.~Atkin and S.~Zohren,
JHEP {\bf 1404} (2014) 118 [{\tt arXiv:1307.3118 [math-ph]}].  

\bibitem{Alday:2009aq}
  L.~F.~Alday, D.~Gaiotto and Y.~Tachikawa,
Lett.\ Math.\ Phys.\  {\bf 91} (2010) 167
[{\tt arXiv:0906.3219 [hep-th]}].  

\bibitem{Nekrasov:2002qd}
  N.~A.~Nekrasov,
Adv.\ Theor.\ Math.\ Phys.\  {\bf 7} (2004) 831
[{\tt hep-th/0206161}].  
  
\bibitem{Nekrasov:2003rj}
  N.~A.~Nekrasov and A.~Okounkov,
{\tt hep-th/0306238}.  

\bibitem{Gamayun:2013auu}
  O.~Gamayun, N.~Iorgov and O.~Lisovyy,
J.\ Phys.\ A {\bf 46} (2013) 335203
[{\tt arXiv:1302.1832 [hep-th]}].  

\bibitem{Jimbo:1979rt}
  M.~Jimbo, T.~Miwa, Y.~M\^{o}ri and M.~Sato,
Physica D {\bf 1} (1980) 80.  

\bibitem{Okamoto}
K.~Okamoto,
Ann. Mat. Pura Appl. {\bf 146} (1987) 337.
      
}

\end{thebibliography}
\end{document}